*materials*

*Review*

# Superfunctional Materials by Ultra-Severe Plastic Deformation

Kaveh Edalati [1,2]

[1] WPI, International Institute for Carbon-Neutral Energy Research (WPI-I2CNER), Kyushu University, Fukuoka 819-0395, Japan; kaveh.edalati@kyudai.jp

[2] Mitsui Chemicals, Inc.—Carbon Neutral Research Center (MCI-CNRC), Kyushu University, Fukuoka 819-0395, Japan

**Abstract:** Superfunctional materials are defined as materials with specific properties being superior to the functions of engineering materials. Numerous studies introduced severe plastic deformation (SPD) as an effective process to improve the functional and mechanical properties of various metallic and non-metallic materials. Moreover, the concept of ultra-SPD—introducing shear strains over 1000 to reduce the thickness of sheared phases to levels comparable to atomic distances—was recently utilized to synthesize novel superfunctional materials. In this article, the application of ultra-SPD for controlling atomic diffusion and phase transformation and synthesizing new materials with superfunctional properties is discussed. The main properties achieved by ultra-SPD include: (i) high-temperature thermal stability in new immiscible age-hardenable aluminum alloys; (ii) room-temperature superplasticity for the first time in magnesium and aluminum alloys; (iii) high strength and high plasticity in nanograined intermetallics; (iv) low elastic modulus and high hardness in biocompatible binary and high-entropy alloys; (v) superconductivity and high strength in the Nb-Ti alloys; (vi) room-temperature hydrogen storage for the first time in magnesium alloys; and (vii) superior photocatalytic hydrogen production, oxygen production, and carbon dioxide conversion on high-entropy oxides and oxynitrides as a new family of photocatalysts.

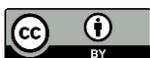

## 1. Introduction to Severe Plastic Deformation

Severe plastic deformation (SPD) is defined as a process in which a large plastic strain is introduced into a material without significant changes in dimensions [1,2]. Several main SPD techniques were invented based on torsion, extrusion, rolling, and forging to achieve high strains without dimensional changes [3,4]. The main effect of SPD is the generation of ultrafine-grained (UFG) microstructure with high-angle grain boundaries and a high density of lattice defects [5,6]. Such microstructural features give superior functional properties to a material that are usually superior to the normal functions of engineering materials [7,8]. Due to these superior functional properties of severely deformation materials, they have sometimes been referred to as superfunctional materials [9]. For example, while low-temperature superplasticity is considered enhanced functionality, room-temperature superplasticity is considered superfunctionality, which usually cannot be achieved by conventional processing routes for many alloys. Another example is hydrogen storage at low temperatures, which is enhanced functionality, while room-temperature hydrogen storage is superfunctionality that cannot be achieved, particularly in Mg-based alloys, by conventional processing routes.

As a brief history of SPD, it can be mentioned that the field appeared first in ancient times for sword making [10], its scientific principles were established in the 1930s [11,12], it became applicable to large samples by extrusion-based techniques in the 1970s [13], and it became a popular process to produce UFG materials following a publication in the 1980s [14]. The progress in the SPD field and its potential to achieve superior properties were discussed in several review papers [1–8], while the most recent developments were extensively discussed in several papers, published in a special issue of Materials Transactions in 2019 [15] and a corresponding overview paper published in 2022 [9]. One of the recent progresses in the field is the application of ultra-SPD as a tool to synthesize new materials [16]. In this article, recent advances in the production of superfunctional materials by ultra-SPD are reviewed.

## 2. Ultra-Severe Plastic Deformation

The term ultra-SPD, which was suggested in 2017 [17], refers to SPD processes in which shear strains over 1000 are applied to a specimen so that the thickness of sheared phases becomes geometrically comparable to atomic distances [16]. The introduction of such large shear strains can lead to the atomic-scale mixing of elements in multiphase materials and the formation of new phases with specific functional properties. Moreover, some studies on ultra-SPD processing of single-phase materials suggested that there may be new deformation stages beyond stage V, which can result in deviations of properties from the apparent steady state [16,18]. The introduction of such large strains, which are sometimes up to 100,000, can be realized mainly using the HPT method, in which a disc [19] or ring [20] specimen is torsionally strained under very high pressures between two Bridgman anvils [11]. Shear strain in the HPT method is calculated using the following equation [12].

$$\gamma = \frac{2\pi r N}{h} \tag{1}$$

where $\gamma$ is the shear strain, $r$ is the distance from the disc/ring center, $N$ is the number of anvil rotations and $h$ is the thickness of the disc/ring. This equation suggests that extremely large strains can be achieved in this method by increasing the number of rotations. If it is assumed that two phases with initial sizes of $d_0$ are subjected to ultra-SPD, their thicknesses are geometrically reduced based on the following equation, provided that the two phases co-deform ideally [17].

$$d = d_0/\gamma \tag{2}$$

This equation, which is valid only at large shear strains, indicates that if the two phases with sizes of 10 μm (a normal size in many materials) are subjected to a shear strain of 10,000, their thickness can be reduced to the subnanometer level. As a result, the two phases can be mixed at the atomic level to form a new phase. A strain of 10,000 is extremely high, but it can be achieved in HPT by increasing the number of rotations to over $N = 100$ in a 20 mm diameter disc with a thickness of 0.6 mm.

In reality, the expected co-deformation by pure shear straining of different phases does not occur, particularly when the hardness and plasticity of the phases are different. Therefore, a higher strain is required for atomic-scale mixing. However, a large density of lattice defects is continuously generated during the process leading to extremely fast lattice diffusion, which is sometimes comparable to pipe or surface diffusion [21–24]. Such a fast atomic diffusion can significantly reduce the shear strain needed for atomic-scale mixing. For example, large amounts of TiFe hydrogen storage material could be produced from titanium and iron powders at ambient temperature after applying a rather low shear strain of 800, indicating the occurrence of ultra-fast diffusion [25]. Figure 1 compares the reported static diffusion coefficients with the dynamic lattice diffusion during ultra-SPD processing via HPT for (a) Al-Cu [24] and (b) Al-Zr [26], indicating that the lattice diffusion during SPD can be comparable to surface diffusion which is orders of magnitude faster than static lattice diffusion. The presence of lattice defects together with the dynamic effects of strain and pressure can also influence the thermodynamics of the system and lead to the formation of new phases even from the immiscible systems [17,27–29].

Currently, ultra-SPD is considered an effective tool to produce a wide range of materials such as binary and ternary alloys and intermetallics [16,17], high-entropy alloys [30], metal hydrides [31,32], and high-entropy ceramics [33]. Table 1 summarizes the results of the application of ultra-SPD to different systems in the author's group and summarizes the main achievements of each system [34–61]. As can be seen in Table 1, the application of ultra-SPD is not limited to the synthesis of conventional metallic alloys, and it has been recently extended to ceramic materials. In the following sections, some main superfunctional properties achieved after ultra-SPD are discussed.

**Table 1.** Materials processed by ultra-SPD to shear strain strains over 1000 and their main properties or features.

| System | Maximum Shear Strain | Properties/Features | Reference |
|---|---|---|---|
| Mg-Li | 7800 | Room-temperature Sperplasticity | [45] |
| $Mg_2X$ (X: 21 elements) | 5500 | Hydrogen storage | [43] |
| Mg-Ti | 5500 | Hydrogen storage | [41] |
| Mg-Zr | 55,000 | Hydrogen storage in new phases | [42] |
| Mg-Hf | 3900 | Biocompatible new phases | [53] |
| Mg-V-Cr | 50,000 | Hydrogen storage | [47] |
| $Mg_4NiPd$ | 59,000 | Room-temperature hydrogen storage | [48] |
| MgTiVCrFe | 12,000 | Hydrogen storage | [50] |
| $MgTiH_4$ | 17,000 | Hydrogen storage | [51] |
| Al-Ca | 39,000 | High-temperature thermal stability | [55] |
| Al-Fe | 39,000 | High-temperature thermal stability | [49] |
| AlNi | 4700 | High hardness | [34] |
| $Al_3Ni$ | 4700 | High hardness | [36,38] |
| Al-Cu | 3900 | Ultra-fast diffusion | [24] |
| Al-Zn | 7800 | Room-temperature superplasticity | [46] |
| Al-Zr | 39,000 | Age hardening and thermal stability | [26] |
| Al-La-Ce | 39,000 | Age hardening and thermal stability | [56] |
| TiAl | 2000 | High strength and high plasticity | [35] |
| TiV | 5500 | Hydrogen storage without activation process | [44] |
| Ti-Nb | 5900 | Biocompatible with high strength and low elastic modulus | [52] |
| TiZrHfNbTa | 2000 | Biocompatible with high strength and low elastic modulus | [57] |
| $TiZHfNbTaO_{11}$ | 7800 | Photocatalytic hydrogen production and $CO_2$ conversion | [54,59] |
| $TiZrHfNbTaO_6N_3$ | 3900 | Photocatalytic hydrogen production and $CO_2$ conversion | [58,60] |
| $TiZrNbTaWO_{12}$ | 3900 | Photocatalytic oxygen production | [61] |
| FeNi | 3900 | Ultra-fast phase transformation | [39] |
| $Ni_2AlTi$ | 4700 | High strength | [37] |
| Nb-Ti | 3900 | Superconductivity | [40] |

*2.1. Thermal Stability in New Immiscible Age-Hardenable Aluminum Alloys*

UFG materials processed by SPD, particularly materials with low melting temperatures such as Al-based alloys, exhibit high hardness, but they usually suffer from poor thermal stability at high temperatures [62–64] or even sometimes at ambient temperature during elongated periods [65–67]. It has been a target of many studies to produce Al-based alloys with high strength and high thermal stability for lightweight applications [68]. The microstructure can be stabilized by second-phase particles of an immiscible element, but due to the immiscibility effect, such alloys show negligible solid solution hardening and poor age hardening [69–72]. Ultra-SPD provides an effective path to achieve supersaturation in various Al-based alloys such as Al-Ca [55], Al-Fe [49], Al-Zr [26], and Al-La-Ce [56]. Such a supersaturation can lead to an enhanced solution hardening effect in these alloys. Moreover, further aging of these supersaturated alloys can result in the precipitation of the second phase particles which not only enhance age hardening but also improve thermal stability.

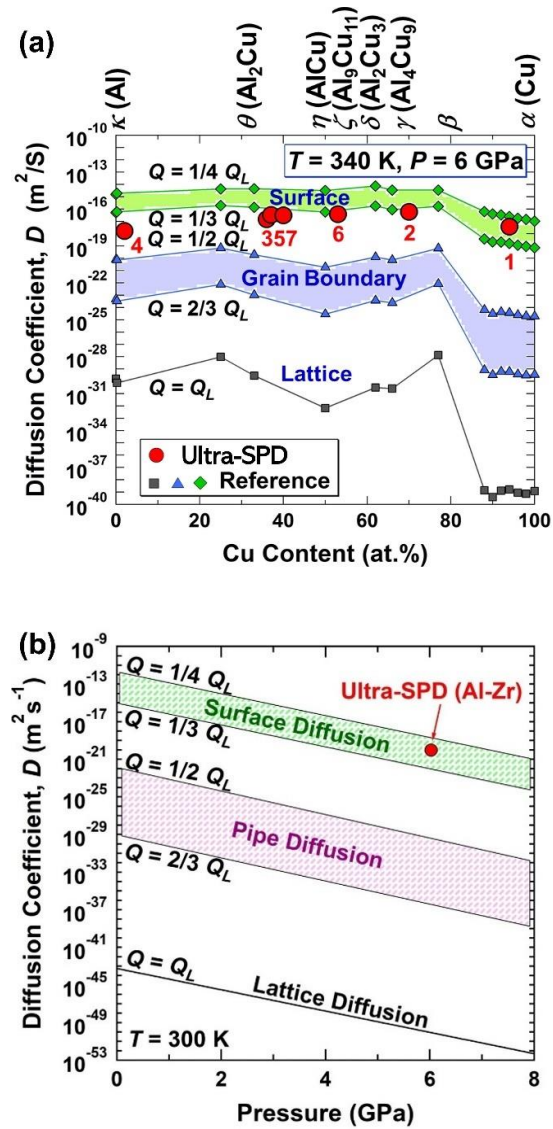

**Figure 1.** Ultra-fast diffusion during ultra-SPD. Estimated diffusion coefficients during ultra-SPD processing via HPT (red circles) for (**a**) Al-Cu alloy plotted against Cu content and (**b**) Al-Zr alloy plotted against pressure in comparison with the reference data calculated using activation energy for lattice diffusion $Q_L$, grain boundary diffusion $(1/2–2/3)Q_L$, and surface diffusion $(1/4–1/3)Q_L$ [24,26].

Figure 2 shows the results achieved using ultra-SPD via the HPT method for an immiscible Al-5Zr (wt%) alloy [26]. The hardness of samples processed with conventional ranges of shear strain is saturated to an apparent steady-state level below 70 Hv, while a significant increase in hardness to about 140 Hv occurs at shear strains over 1000, as shown in Figure 2a. Further heating of the sample to up to 523 K results in age hardening and a further increase of hardness, while the hardness and electrical conductivity of the material remain well stable up to 583 K, as shown in Figure 2b. Moreover, Figure 2b shows that the hardness of ultra-SPD-processed Al-Zr is higher than all data reported so far for the Al-Zr-based alloys even after SPD processing. An examination of the microstructure of this Al-Zr alloy by high-resolution transmission electron microscopy, shown in Figure 2c–f, confirms that such a high hardness and good thermal stability is not only due to the presence of thermodynamically stable Al$_3$Zr intermetallics with the tetragonal structure but also due to the precipitation of orthorhombic AlZr intermetallics at grain boundaries and coherent cubic Al₃Zr intermetallics in grain interiors [26].

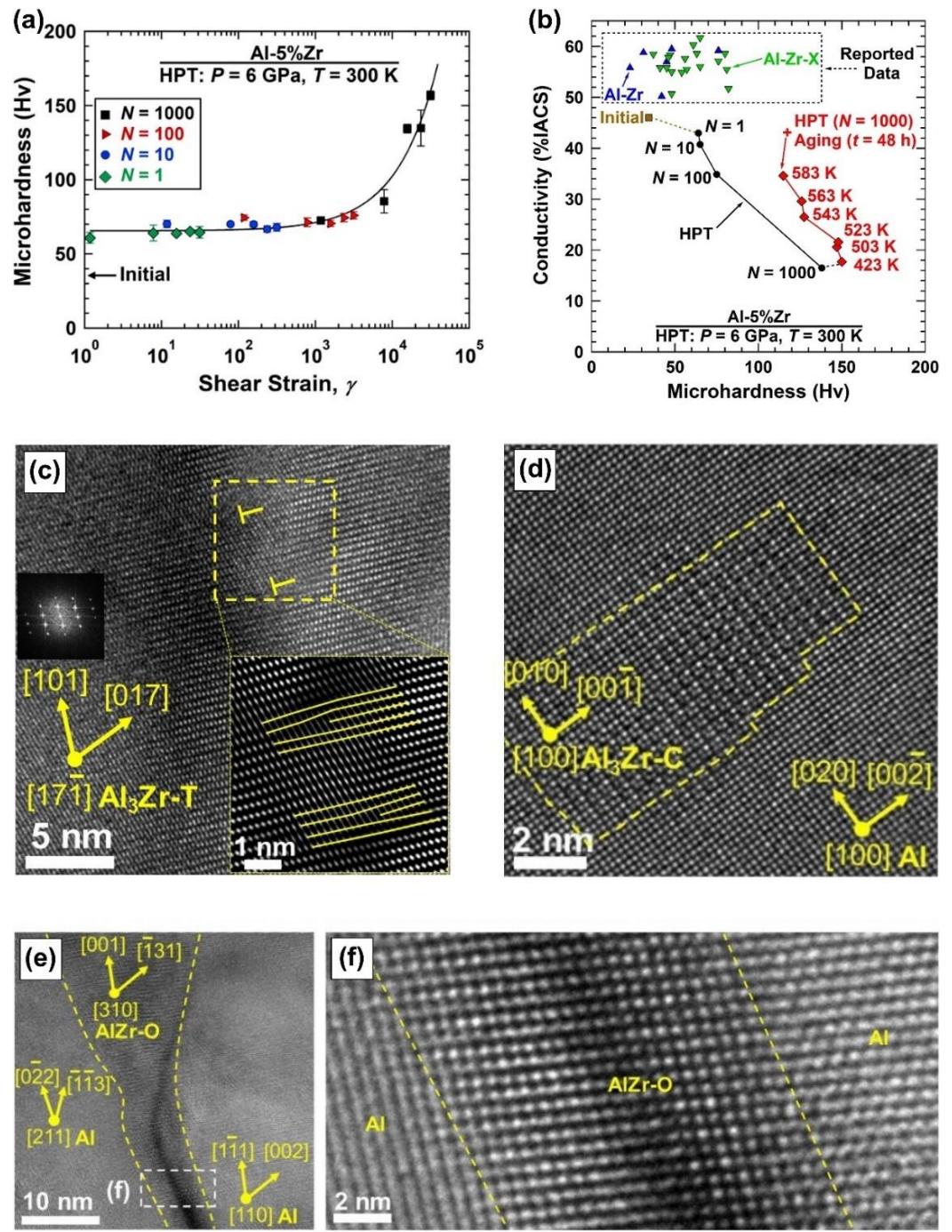

**Figure 2.** Age hardening and high thermal stability in immiscible Al-Zr alloy after ultra-SPD. (**a**) Hardness against shear strain after ultra-SPD through various HPT rotations. (**b**) Electrical conductivity versus hardness after ultra-SPD and aging for 48 h at different temperatures compared to data reported for conventional SPD-processed binary and ternary Al-Zr alloys. Lattice images of (**c**) Al₃Zr intermetallic with tetragonal structure, (**d**) Al₃Zr with cubic structure precipitated in aluminum matrix after aging, and (**e**,**f**) AlZr with orthorhombic structure formed at aluminum grain boundaries, where (**f**) is magnified view of rectangular region in (**e**) [26].

## 2.2. Room-Temperature Superplasticity in Magnesium and Aluminum Alloys

The nanostructured materials show high strength, but their plasticity is usually weak at low homologous temperatures due to the limited activity of dislocations [73,74]. However, the plasticity of nanograined materials can be enhanced and even reach the superplastic level (over 400% elongation) at homologous temperatures over 0.5 due to enhanced atomic diffusion and activation of thermal deformation mechanisms such as grain-boundary sliding [75,76]. This fact was used in the SPD community to achieve enhanced high-temperature superplasticity from the 1980s [14] to nowadays [77]. However, one can expect that if grain-boundary sliding becomes the dominant deformation mechanism at low homologous temperatures, it would be possible to achieve superplasticity even at room temperature.

Grain boundary sliding in superplasticity is described by a relation of strain rate versus stress ($\dot{\varepsilon} - \sigma$) for a material having a Burgers vector of $b$, a shear modulus of $G$, and a mean grain size of $d$ [76].

$$\dot{\varepsilon} = \frac{AD Gb}{kT}\left(\frac{b}{d}\right)^p \left(\frac{\sigma}{G}\right)^n \quad (3)$$

where $A$ is a constant close to 10, $k$ is Boltzmann's constant, $p$ is the exponent of inverse grain size which is usually 2 for grain boundary sliding, $n = 1/m$ is the exponent of stress where $m$ is the strain-rate sensitivity, and usually has a value of $m = 0.5$ in superplastic deformation, and $D$ is the grain boundary diffusion coefficient. Based on this equation, in addition to grain size reduction, which can enhance the grain-boundary sliding, increasing the grain boundary diffusion through grain boundary engineering is another key factor to achieve room-temperature superplasticity. The concept of ultra-SPD was employed for such grain boundary engineering and achieving room-temperature superplasticity for the first time in magnesium and aluminum alloy at low homologous temperatures of 0.37 and 0.36, respectively [45,46].

The ultra-SPD-processed alloys Mg-Li and Al-Zn showed 440% and 480% elongation at room temperature, respectively, as shown in Figure 3a for the Al-Zn alloy [46]. Both alloys had two phases, including Li- and Zn-rich phases with fast grain boundary diffusion and Mg- and Al-rich phases with slow grain boundary diffusion. Ultra-SPD significantly enhanced the fraction of interphase boundaries that have fast boundary diffusion. It also led to the segregation of Li in the boundaries of Mg-rich/Mg-rich grains and Zn in the boundaries of Al-rich/Al-rich grains, as shown in Figure 3b–e [46]. Such segregation enhanced the grain boundary diffusion and made it comparable to those of interphase boundaries, and accordingly, resulted in room-temperature superplasticity. These results are of significance because none of the earlier attempts by different processing routes including SPD were successful to achieve room-temperature superplasticity in Mg-Li, Al-Zn, or any other Mg-based and Al-based alloys [77–80].

### 2.3. High Strength and High Plasticity in Nangrained Intermetallics

Intermetallics show a variety of functional properties because they have features of both metals and ceramics. Nanostructuring usually enhances the properties of these materials, and thus, there have been numerous attempts to produce nanostructured intermetallics [81–84]. The synthesis of nanopowders by chemical methods, ball milling, and gas condensation, are some popular methods, but these nanopowders need to be consolidated at high temperatures when a bulk sample is needed, and this can cause grain coarsening [81–84]. The application of SPD to coarse-grained intermetallics, synthesized by melting techniques, is another effective two-step strategy to produce bulk nanostructured samples [85–88]. Ultra-SPD can be used as a single-step process to synthesize ultra-hard bulk nanostructured intermetallics directly from the elemental powders. Several intermetallics such as various Mg-based intermetallics [43], AlNi [34], Al$_3$Ni [36,38], TiAl [35], FeNi with the L1$_0$ structure [39] and Ni$_2$AlTi [37] were synthesized in recent years by ultra-SPD or ultra-SPD followed by low-temperature annealing.

For a successful synthesis of hard intermetallics, in addition to ultra-high shear strain, high pressure is also needed. HPT is the main method that can be used for such applications because HPT applies to any kind of materials including hard metals [89,90], intermetallics [85,86], and even glasses [91,92]. It should be noted that the application of ultra-SPD can sometimes lead to the formation of new intermetallics, which do not exist in the equilibrium phase diagram such as Mg$_4$NiPd with the CsCl-type cubic structure [48]. In some systems with complete immiscibility even in the liquid form such as Mg-Zr [42] and Mg-Hf [53], new phases were discovered after ultra-SPD. The speed of synthesis of intermetallics by ultra-SPD is usually very fast and some phases such as L1$_0$ FeNi, which forms only within an astronomical time scale, can be formed within a few hours by ultra-SPD [39].

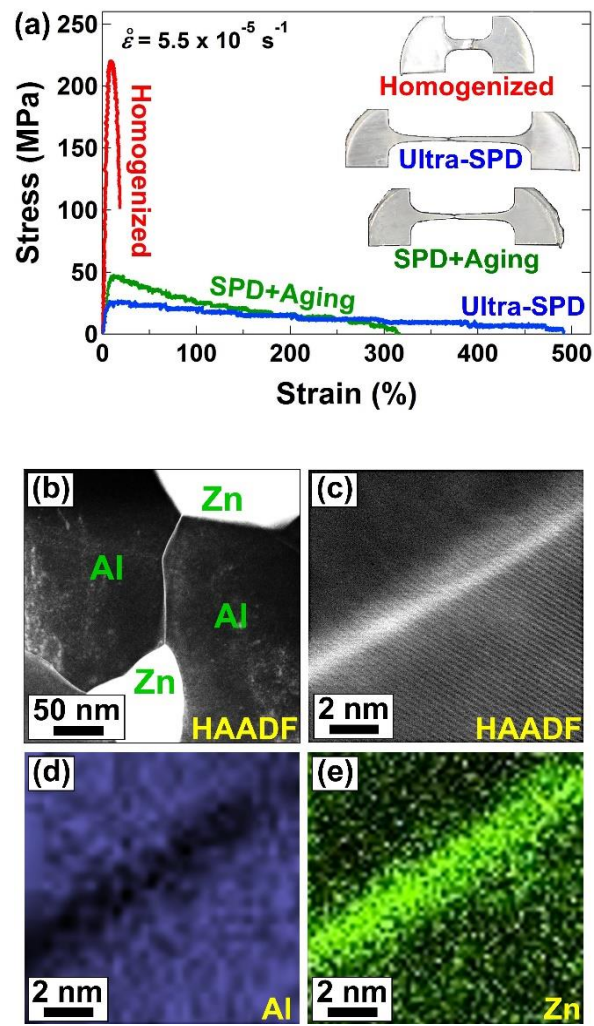

**Figure 3.** Room-temperature superplasticity in Al-Zn alloy after ultra-SPD. (**a**) Tensile stress-strain curves, including appearance of tensile specimens after pulling to failure, after homogenization, ultra-SPD and ultra-SPD followed by 100-day aging at room temperature. (**b**) High-angle annular dark-field (HAADF) image showing the distribution of Al and Zn atoms after ultra-SPD processing in which bright and dark contrasts correspond to Al and Zn, respectively. (**c**) HAADF lattice image of Al-Al grain boundary and corresponding elemental mapping with (**d**) Al and (**e**) Zn after processing by ultra-SPD via HPT [46].

In addition to ultrahigh hardness, the ultra-SPD-processed intermetallics can sometimes exhibit unusually high plasticity [35]. Figure 4a compares the yield strength and plasticity of the TiAl intermetallic synthesized by ultra-SPD with those reported for coarse-grained and nanograined TiAl-based intermetallics. The ultra-SPD-processed TiAl shows an excellent combination of strength and plasticity, which is beyond the trade-off strength-plasticity relationship expected for the TiAl-based intermetallics. Such a good combination of strength and ductility was attributed to the bimodal microstructure of the intermetallic, the presence of many nanotwins shown in Figure 4b, and the activation of different deformation mechanisms under compression such as twining, dislocation slip, and grain boundary sliding, as shown in Figure 4c,d [35].

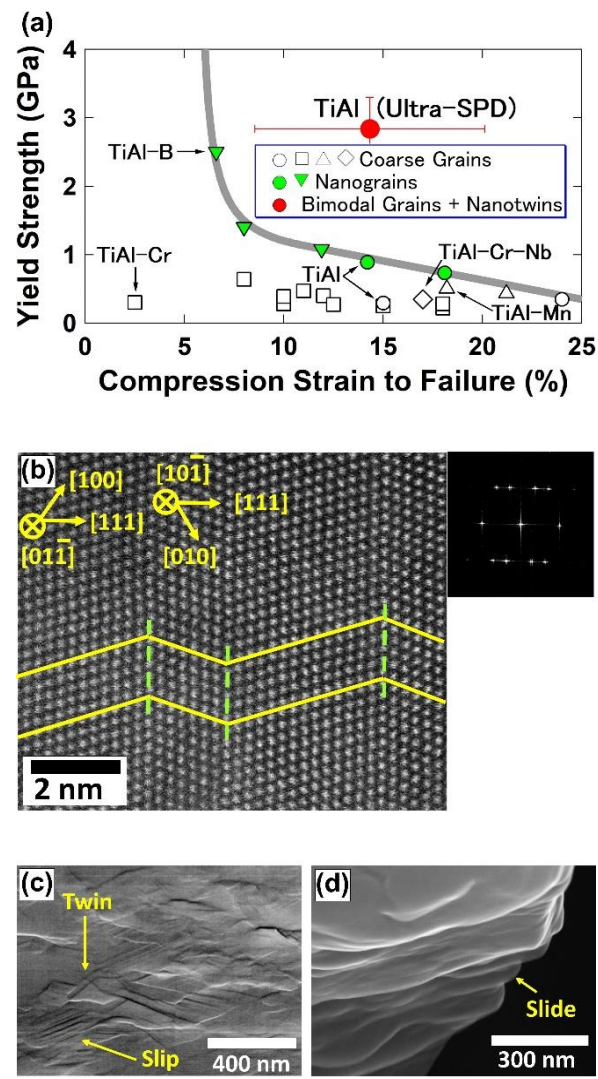

**Figure 4.** High strength and high plasticity in TiAl intermetallic after ultra-SPD. (**a**) Yield strength versus compression strain to failure for TiAl synthesized by ultra-SPD and low-temperature annealing in comparison with reported data for coarse-grained and nanograined TiAl alloys. (**b**) Formation of nanotwins in TiAl after ultra-SPD and annealing. (**c**,**d**) Activation of different deformation mechanisms observed by scanning electron microscopy on surface of TiAl specimen after compression test [35].

*2.4. Low Elastic Modulus and High Strength in Biocompatible Binary and High-Entropy Alloys*

In biomaterials used for orthopedic implant applications, in addition to biocompatibility and high strength, the elastic modulus is an important property that should be as close as possible to the elastic modulus of human bone (10–30 GPa) to avoid stress shielding effect [93]. To achieve high strength usually titanium is alloyed with other elements, but such alloying may negatively influence biocompatibility and may not improve the elastic modulus. For example, the elastic modulus for titanium, Ti-6Al-4V, and Ti-6Al-7Nb (wt.%) are 105 GPa, 110 GPa, and 105 GPa, respectively, while aluminum and vanadium are considered toxic in long-term use of implants [93]. Processing by SPD can lead to increasing the strength of titanium without the addition of alloying elements, while it was shown that the biocompatibility of nanostructured titanium is also better than the coarse-grained metal [94–96]. This fact has led to the commercialization of severely deformed titanium for implant applications [97].

Ultra-SPD was recently used to achieve some of the highest hardness values ever reported for biomaterials by mechanical alloying of titanium with biocompatible elements such as niobium, zirconium, tantalum, and hafnium [52,57]. The applicability of ultra-SPD to almost any kind of system with a minor contamination effect allows synthesizing of bulk alloys with the best combination of high hardness and low elastic modulus [52,57]. Figure 5a shows the variation of elastic modulus against the niobium content in the Ti-Nb alloys synthesized by ultra-SPD. It is evident that a Ti-25Nb alloy (at%) can exhibit an elastic modulus comparable to human bone, while its hardness is as high as 370 Hv [52]. Further alloying to form ternary alloy TiNbZr, medium-entropy alloy TiNbZrTa, and high-entropy alloy TiNbZrTaHf lead to higher hardness values up to 565 Hv with moderate elastic modulus values close to 80 GPa, which are smaller than the elastic modulus of pure titanium [57]. A comparison of these findings with the reported data in the literature in Figure 5b confirms that the biomaterials synthesized by ultra-SPD show some of the best combinations of high hardness and low elastic modulus without the addition of toxic elements such as aluminum and vanadium [52,57].

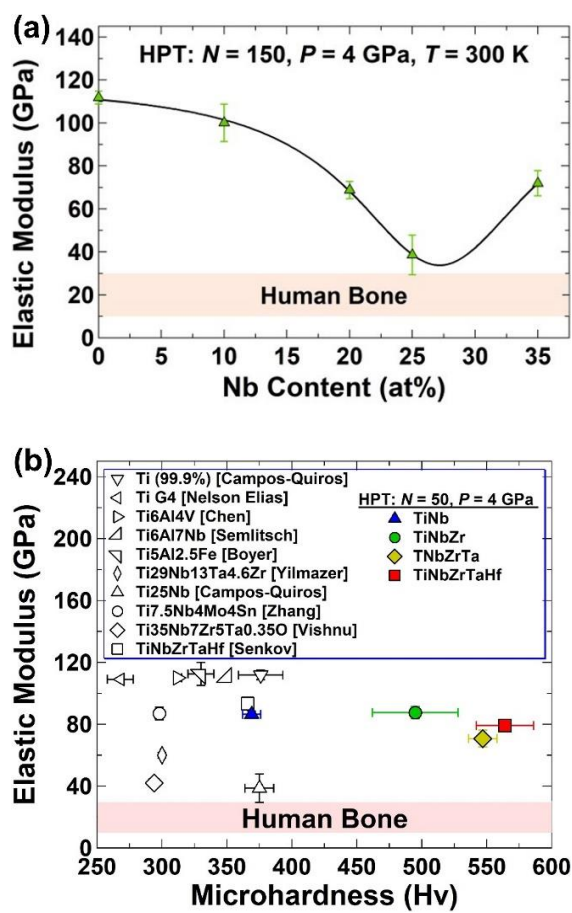

**Figure 5.** Low elastic modulus and high hardness of biocompatible alloys after ultra-SPD. (**a**) Elastic modulus versus Nb content in Ti-Nb biomaterial synthesized by ultra-SPD in comparison with elastic modulus of human bone [52]. (**b**) Elastic modulus versus microhardness for binary alloy TiNb, ternary alloy TiNbZr, medium-entropy alloy TiNbZrTa and high-entropy alloy TiNbZrTaHf synthesized by ultra-SPD in comparison with data reported for some alloys by other synthesis or SPD methods [57].

## 2.5. Superconductivity and High Strength in Nb-Ti Alloys

The Nb-Ti alloys are the most widely used superconductors in the industry. In the Nb-Ti alloys, which are mainly used in superconducting magnets, the presence of titanium nanoparticles pins vortices, and enhances the critical current density in a magnetic field [98–101]. The Nb-Ti alloys are mainly fabricated by metal forming techniques such as wire drawing and extrusion or rolling followed by long-time annealing and this process should be repeated many times to have a good distribution of titanium nanoparticles in the niobium matrix [98–101]. However, ultra-SPD can provide a one- or two-step process to synthesize the Nb-Ti alloys with superconducting properties comparable to or even better than those achieved by repeated metal forming and annealing [40].

To fabricate the Nb-Ti superconductors by ultra-SPD, a mixture of titanium and niobium is first mechanically alloyed by ultra-SPD to achieve a supersaturated solid solution and then annealed for an appropriate time to have the formation of titanium nanograins, as illustrated in Figure 6a [40]. As shown in Figure 6b, the Nb-Ti alloy processed by ultra-SPD exhibits superconducting properties below a critical temperature of 8.9 K, which is comparable with the reported properties for industrial Nb-Ti superconductors [98–101]. However, an advantage of ultra-SPD is that it is a single- or double-step synthesis process, while commercial Nb-Ti superconductors are fabricated by repeated cold working and annealing. Moreover, the ultra-SPD-processed Nb-Ti shows higher tensile strength, bending strength, and hardness compared to commercial superconductors, as shown in Figure 6c. In conclusion, ultra-SPD can be effectively used to synthesize nanostructured superconductors, and this can be of significance because nanostructuring can also enhance the critical temperature for superconductivity due to the quantum size effect [102].

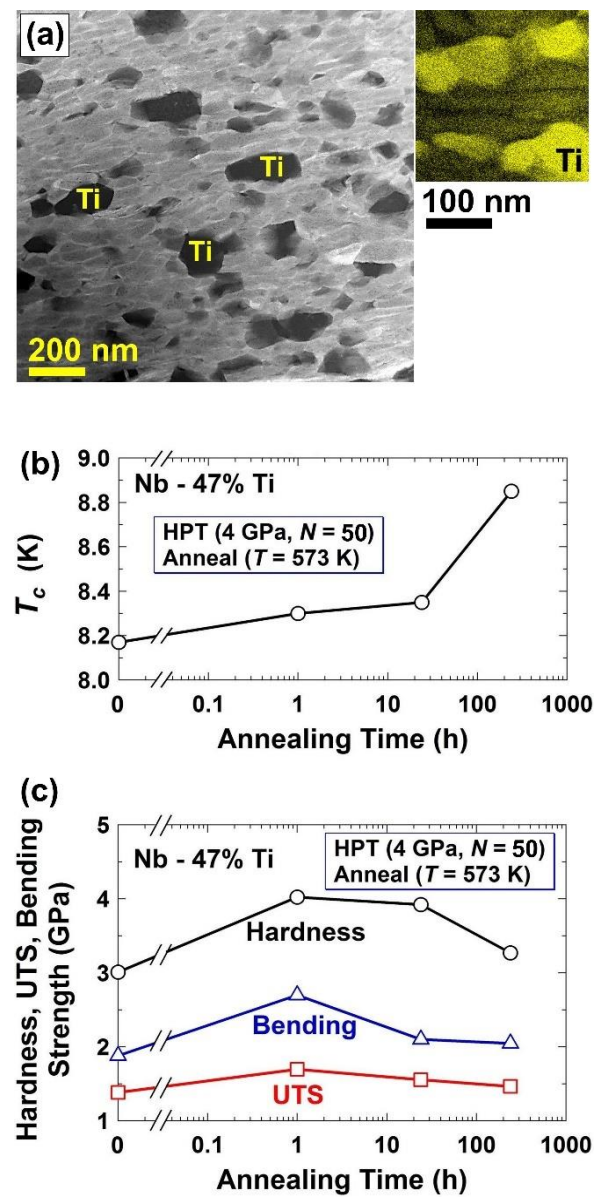

**Figure 6.** Synthesis of superconductors with high strength by ultra-SPD. (**a**) Microstructure of Nb-Ti synthesized by ultra-SPD followed by 10 days annealing at 573 K, examined by bright-field imaging (**left**) and elemental mapping (**right**) in scanning-transmission electron microscopy. (**b**) Critical temperature for superconductivity and (**c**) ultimate tensile strength, bending strength and hardness for Nb-Ti superconductor synthesized by ultra-SPD and annealed at 573 K for various periods of time [40].

*2.6. Room-Temperature Hydrogen Storage in Magnesium Alloys*

In the realization of hydrogen as a clean fuel, storage of hydrogen in a safe and high-density form is still a challenging task [103]. Solid-sate hydrogen storage in the form of metal hydrides is a promising technology because of the compact and low-pressure storage of hydrogen [104,105]. Among various materials for hydrogen storage in the form of hydrides, magnesium and its alloys are the first and most investigated due to their large storage capacity, low price, and high abundance of magnesium on the Earth's crust [106,107]. However, magnesium and its alloys suffer from high dehydrogenation temperatures due to the high thermodynamic stability of their hydrides and slow kinetics [106,107]. The SPD process including equal-channel angular pressing [108–110] and HPT [111–113] has received high attention from different research groups in the past two decades because the process can successfully solve the kinetic drawbacks of $MgH_2$ and Mg-based hydrogen storage materials. Moreover, some studies recognized the potential of ultra-SPD in synthesizing various Mg-based hydrogen storage materials [31,32], including $Mg_2X$ intermetallics (X: transition metals) [43], Mg-Ti [41,51], Mg-Zr [42], Mg-Hf [53], Mg-V-Cr [47] and MgVTiCrFe [90], although not all these alloys practically exhibited hydrogen uptake within the experimental conditions examined. Despite all these attempts, the synthesis of materials with appropriate thermodynamics that can reversibly store hydrogen at room temperature is still a challenging task [106,107].

A combination of hydrogen binding energy engineering and ultra-SPD was successfully employed in recent years to produce the first Mg-based alloy with reversible hydrogen storage capability at room temperature [48]. As shown in Figure 7a using first-principles calculation, a hydrogen binding energy close to -0.1eV per hydrogen atom is desirable for room-temperature hydrogen storage. It was suggested that such binding energy is theoretically achievable in the 4Mg-1N-Pd octahedral sites of a $Mg_4NiPd$ alloy with a CsCl-type cubic structure. The fabrication of this alloy by a melting method resulted in the undesirable formation of three phases of $Mg_8Ni_3Pd$, $MgNi_2$, and $Mg_5Pd_2$, but the application of ultra-SPD resulted in the homogenization of three elements in the form of a single CsCl-type cubic phase as illustrated in Figure 7b. This phase exhibited reversible room-temperature hydrogen storage at ambient temperature for at least five cycles, as shown in Figure 7c, while it kept its CsCl-type crystal structure after hydrogenation/dehydrogenation cycling as shown in Figure 7d. $Mg_4NiPd$ synthesized by ultra-SPD can be considered the first Mg-based alloy with room-temperature hydrogen storage capability, although its storage capacity (0.9 wt% in Figure 7c) is not as high as high-temperature hydrogen storage materials such as pure magnesium (7.6 wt%) [103–107]. These results suggest that a combination of theoretical calculations through binding-energy engineering and ultra-SPD can be effectively used to develop hydrogen storage materials with promising properties for room-temperature hydrogen storage [48].

*2.7. Photocatalytic Water Splitting and $CO_2$ Conversion on High-Entropy Ceramics*

$CO_2$ emission from the utilization of fossil fuels has resulted in global warming as one of the major crises of the 21st century [114]. The usage of hydrogen as a zero-$CO_2$ emission fuel and the conversion of $CO_2$ to hydrocarbons or reactive gases, such as CO, are two major solutions to address the $CO_2$ emission issues [115,116]. Photocatalytic water splitting to hydrogen [117] and $CO_2$ conversion [118] are perhaps the cleanest technologies in this regard, but the efficiency of photocatalysis is still low for practical applications. Therefore, there are significant attempts all over the world to discover novel photocatalysts with high activity. One of these attempts is the use of the concept of ultra-SPD to introduce highly active and stable photocatalysts [54,58–61].

Ultra-SPD followed by oxidation and/or nitriding was used to introduce the first high-entropy photocatalysts containing at least five principal cations together with oxygen and nitrogen anions [54,58–61]. The high-entropy oxide $TiZrHfNbTaO_{11}$ [54,59] and the high-entropy oxynitride $TiZrHfNbTaO_6N_3$ [58,60] showed photocatalytic activity for hydrogen production and $CO_2$ conversion under UV light, while the high-entropy oxide $TiZrNbTaWO_{12}$ [61] exhibited photocatalytic activity for oxygen production under visible light. Although oxynitrides usually suffer from poor stability, the high-entropy oxynitride photocatalyst showed good cyclic stability for hydrogen production even after six-month storage, as shown in Figure 8a [58]. Moreover, TiZrHfNbTaO$_{11}$ shows high activity for $CO_2$ conversion comparable to benchmark P25 TiO$_2$ photocatalyst, while TiZrHfNbTaO$_6$N$_3$ shows the highest photocatalytic activity (per catalyst surface area) reported so far for $CO_2$ to CO conversion, as shown in Figure 8b [60]. The high photocatalytic activity of this new family of photocatalysts is not only due to high light absorbance, as shown in Figure 8c, but also due to high $CO_2$ physical adsorption and chemisorption, as shown in Figure 8d using diffuse reflectance infrared Fourier transform spectroscopy [60]. It should be noted the high stability observed in high-entropy photocatalysts is a general feature of high-entropy alloys [119,120] and ceramics [33,121]. Taken altogether, these studies confirm the high potential of ultra-SPD in discovering novel materials for photocatalysis. Although the HPT method is the most potent method to introduce significant shear strain for the synthesis of these new materials, some continuous methods such as accumulative roll-bonding are basically applicable for ultra-SPD processing, provided that very large numbers of passes can be conducted in these methods [122,123].

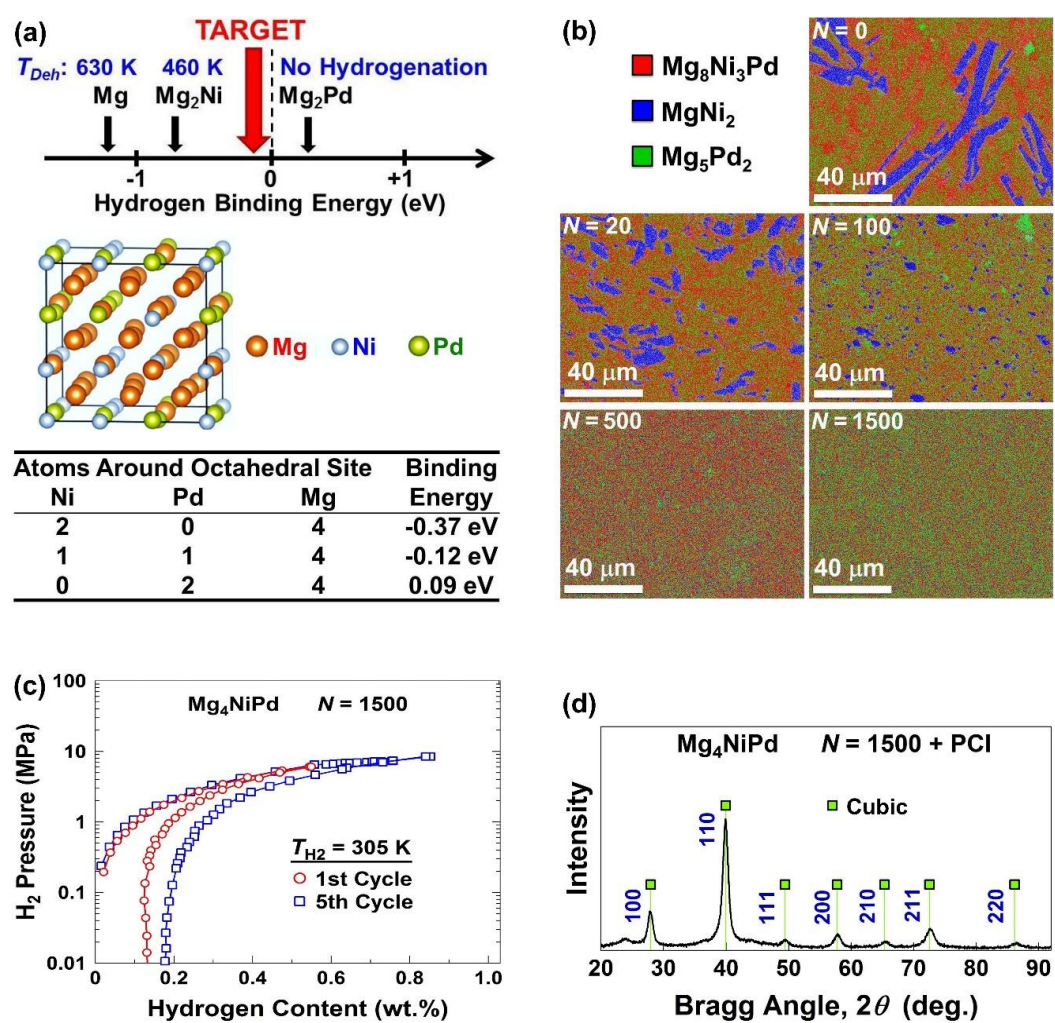

**Figure 7.** Hydrogen storage at room temperature in magnesium alloy designed by binding-energy engineering any synthesized by ultra-SPD. (**a**) Concept of binding-energy engineering used to design Mg$_4$NiPd with appropriate binding energy at 1Ni-1Pd-4Mg octahedral sites. (**b**) Elemental mapping of Mg$_4$NiPd before ($N$ = 0) and after HPT processing for $N$ = 20, 100, 500 and 1500 rotations. (**c**) Pressure-composition isotherms and (**d**) X-ray diffraction profile after five-cycle hydrogenation/dehydrogenation for Mg$_4$NiPd synthesized by ultra-SPD via 1500 HPT turns [48].

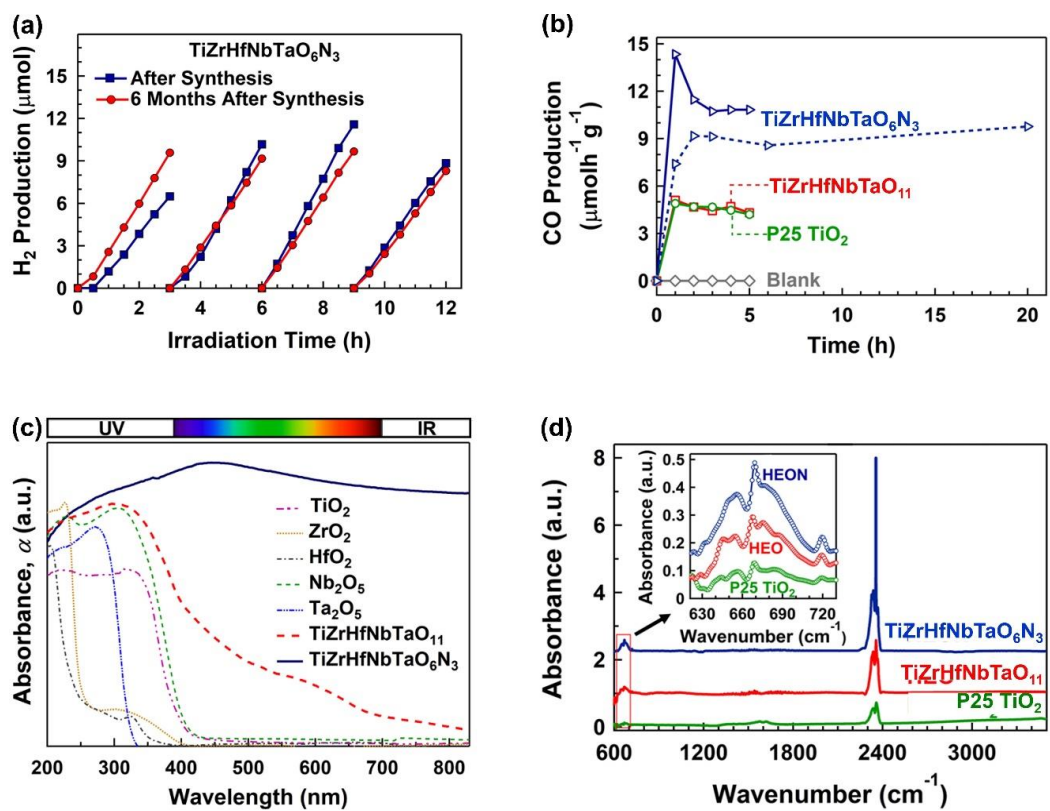

**Figure 8.** Application ultra-SPD for developing high-entropy oxides and oxynitrides as a new family of photocatalysts for photocatalytic hydrogen production and $CO_2$ conversion. (**a**) Photocatalytic hydrogen production versus light irradiation time on $TiZrHfNbTaO_3N_6$ synthesized by ultra-SPD [58]. (**b**) Photocatalytic $CO_2$ to CO conversion versus light irradiation time on $TiZrHfNbTaO_{11}$ and $TiZrHfNbTaO_3N_6$ synthesized by ultra-SPD in comparison with P25 $TiO_2$ benchmark photocatalyst [60]. (**c**) Light absorbance in $TiZrHfNbTaO_{11}$ and $TiZrHfNbTaO_3N_6$ in comparison with binary oxides [58]. (**d**) Diffuse reflectance infrared Fourier transform spectra for $TiZrHfNbTaO_{11}$ and $TiZrHfNbTaO_3N_6$ in comparison with P25 $TiO_2$ where intensities of spectra at 665 cm$^{-1}$ and 2350 cm$^{-1}$ represent chemisorption and physisorption of $CO_2$ on catalyst surface [60].

## 3. Concluding Remarks and Future Outlook

Superfunctional materials, with properties superior to the normal functions of engineering materials, can be readily synthesized by ultra-severe plastic deformation (ultra-SPD). In ultra-SPD, the shear strain is significantly increased to over 1000 so that the thicknesses of sheared phases become comparable to atomic distances, and accordingly, atomic-scale mixing of phases occurs. The application of ultra-SPD has resulted in various superfunctional properties such as superior thermal stability, room-temperature superplasticity, high strength/plasticity, low elastic modulus combined with high strength and biocompatibility, superconductivity, room-temperature hydrogen storage, and superior photocatalytic water splitting and $CO_2$ conversion. These findings not only have expanded the applications of SPD but also successfully introduced new families of metallic, intermetallic, composite, and ceramic materials with superfunctional properties. The application of ultra-SPD, at least on the laboratory scale, is expected to grow more significantly in the future particularly with the global need to discover new energy materials for carbon-neutral energy applications. However, the field should be empowered by close connections with theoretical studies and numerical simulation for efficient material design.

**Funding:** The author was supported in part by the Light Metals Educational Foundation of Japan, in part by the MEXT, Japan through Grants-in-Aid for Scientific Research on Innovative Areas (JP19H05176 & JP21H00150), and in part by the MEXT, Japan through Grant-in-Aid for Challenging Research Exploratory (JP22K18737).

**Institutional Review Board Statement:**

**Informed Consent Statement:**

**Data Availability Statement:**

**Conflicts of Interest:** The author declares no conflicts of interest.